\documentstyle[aps,prb,preprint]{revtex}

\begin{document}
\author{{\bf Miodrag} {\bf L.} {\bf Kuli\'{c} }}
\address{Physikalisches Institut, Theorie III, Universit\"{a}t Bayreuth, \\
95440 Bayreuth, Germany}
\title{{\bf Thermoelectric Effects in S-N-S Weak Links with Heavy Fermions as The
Normal Metal: A Possibility for Thermosensors}}
\date{18 November 2000}
\maketitle

\begin{abstract}
It is shown that S-N-S weak links with an heavy-fermion metal as the weak
link (N) can be useful thermosensors. This property is due to the large
thermopower of heavy-fermion metals which is of the order $10^{-6}$ $V/K$.
The longitudinal sound can also generate voltage oscillation due to the
large Gr\"{u}neisen parameter, $\Omega _{hf}\sim 10^{2}$, in heavy-fermions.
Similar effects are expected with other Kondo systems as the weak link.
\end{abstract}

\newpage

\section{\ Introduction}

It is well known, that below some characteristic temperature $T^{\ast
}(\approx 20-100$ $K)$ heavy-fermion (HF) metals , which contain magnetic
ions with 4f or 5f electrons, are characterized by the large quasiparticle
mass $m_{hf}$. The latter is several hundred times larger than the electron
mass $m_{e}$ - see for instance \cite{Heavy1}, and these systems are
strongly renormalized Fermi liquids below $T^{\ast }$. As a consequence,
various thermodynamic and transport properties are significantly
renormalized below $T^{\ast }$. For instance, there is a significant
increase : (1) in the low temperature specific heat ($C_{p}=\gamma _{hf}T$),
where $\gamma _{hf}\sim 10^{3}\gamma $ with the typical value for alkali
metals$\ \gamma \sim 1mJ$ $mol^{-1}K^{-2}$; (2) in the spin paramagnetic
susceptibility; (3) in the electronic Gr\"{u}neisen parameter $\Omega
_{hf}=-d\ln T^{\ast }/du$ ($u$ is the strain field), where for instance $%
\Omega _{hf}$ of the HF metal $UPt_{3}$ is 100 times larger than in standard
alkali metals; (4) in the thermoelectric power $Q_{hf}\sim 10^{2}$ $Q$ where 
$Q$ is the value for the standard alkali metals. The main reason for this
renormalization is the presence of the Abrikosov-Suhl-Kondo (ASK) resonance
in the quasiparticle density of states at the Fermi level. The latter is due
to the Kondo effect in the lattice of 4f (or 5f) magnetic ions - for more
details see \cite{Heavy1}, \cite{Heavy2}.

The large longitudinal sound absorption measured \cite{Heavy3} in $UPt_{3}$
around $T\approx 10$ $K$ is explained in Ref. \cite{Kulic} by the large
values of $\Omega _{hf}$ and $Q_{hf}$, which cause large temperature
fluctuations during sound propagation. Note that in standard alkali metals
(which are weakly renormalized Fermi liquids) the effect of temperature
fluctuations on the sound absorption is negligible. In Ref. \cite{KuKe} it
was proposed a setup which might prove the existence of thermoelectric
effects in superconductors. It is also based on the large value of the
thermopower 
\begin{equation}
Q_{hf}\approx -\frac{3T}{\mid e\mid T^{\ast }}  \label{Q}
\end{equation}
in the normal state of the HF metal $UPt_{3}$ ( and also in the
superconducting state near $T_{c}<1$ $K$). It is worth mentioning that in $%
UPt_{3}$ the characteristic value is $\mid Q_{hf}\mid \sim 10^{-6\text{ }}T$ 
$V/K$, where $T$ is measured in $K$. If the HF superconductor with large $%
\mid Q_{hf}\mid $ is part of a ring with two metallic contacts at different
temperatures such a system gives rise to a large induced magnetic flux $%
\Delta \Phi $ inside the hole of the ring. In the ring with an HF
superconductor the quantity $\Delta \Phi /\Phi _{0}T$ is two orders of
magnitude larger than in absence of the HF metal.

In what follows we propose a S-N-S system in which the weak link (N) is made
of an heavy-fermion metal with large thermopower $Q_{hf}$. It is
demonstrated below that such a system is potentially useful for making very
sensitive thermosensors. The sandwich geometry is assumed in which the N-S
contacts lay in the $x-y$ plane, while the $z$-axis is perpendicular to the
contact. Let us assume that the parameters of the system are such that the
superconducting current in the S-N-S weak link is described by the Josephson
current relation, where $I_{S}=I_{c}\sin \varphi $ with $\varphi =\varphi
_{L}-\varphi _{R}$ and $\varphi _{L(R)}$ is the phase of the left (right)
superconducting bank. (The discussion of the latter condition is postponed
to the end.) If the temperature difference, $\Delta T\equiv T_{L}-T_{R}$, is
applied at the contacts between the normal (HF) metal and the left and right
superconducting bank one can show that the phase $\varphi $ fulfills the
equation 
\begin{equation}
\frac{\partial ^{2}\varphi }{\partial t^{2}}-c_{0}^{2}(\frac{\partial
^{2}\varphi }{\partial x^{2}}+\frac{\partial ^{2}\varphi }{\partial y^{2}})+%
\frac{1}{\tau }\frac{\partial \varphi }{\partial t}+\omega _{0}^{2}(\sin
\varphi -\frac{Q_{hf}\Delta T}{I_{c}R_{N}})=0.  \label{2}
\end{equation}

Here, $c_{0}$ is the Swihart velocity, $\omega _{0}(\equiv c_{0}/\lambda
_{J})=\sqrt{2\mid e\mid I_{c}/\hbar C}$ is the Josephson plasma frequency, $%
I_{c}$ is the critical current of the contact, $C$ is the capacity of the
contact, $\tau =R_{N}C$, $R_{N}$ is the resistance of the contact, and $%
\lambda _{J}$ is the Josephson penetration depth. From $Eq.(\label{2}2)$ it
is seen that the term $Q_{hf}\Delta T/R_{N}$ acts as a fixed external
current $I$ in the standard Josephson contact \cite{Aronov}. Based on this
analogy a number of physical phenomena based on $Eq.(2)$ can be studied \cite
{Kulic2} but in this short note we analyze a few of them only.

{\it Thermosensors.} If the temperature difference fulfills the condition $%
\Delta T>\Delta T_{c}=I_{c}R_{N}/Q_{hf}$ the voltage on the contact
oscillates with the frequency (related to the average voltage $<V>=\hbar
\omega /2\mid e\mid $) 
\begin{equation}
\omega =\frac{2\mid e\mid }{\hbar }R_{N}I_{c}\sqrt{(\Delta T/\Delta
T_{c})^{2}-1}.  \label{omega}
\end{equation}
This relation, which is an analogon of the nonstationary Josephson effect,
was first derived in Ref. \cite{Aronov}.

The peculiarity of the HF system is that the threshold temperature
difference is very small, and since $\Delta T_{c,hf}\sim Q_{hf}^{-1}$ it is
practically 100 times smaller than for standard metals. For instance if the
system operates at $T=2$ K one has $Q_{hf}\sim 10^{-6}$ $V/K$ and if the HF
metal is characterized by the small value of $I_{c}R_{N}=V_{c}\sim
(10^{-10}-10^{-13})$ $V$ then $\Delta T_{c,hf}\sim (10^{-4}-10^{-7})$ $K$.
As a comparison, by using as the weak link (N) a standard metal with $Q\sim
10^{-8}$ $V/K$ (at the same temperature) one obtains $\Delta T_{c}\sim
(10^{-2}-10^{-5})$ $K$. The latter value is at least by two orders of
magnitude larger than in the case of an HF metal. From this analysis one
concludes, that the small value of $\Delta T_{c,hf}$ opens a possibility of 
{\it measuring very small temperature difference}. The latter can be done by
measuring the average voltage on the junction, or by measuring the frequency
of the emitted radiation. On such a ground one can make {\it thermosensors}.
One expects also that by applying low-frequency radiation on the contact one
can generate small temperature difference in it \cite{Kulic2}.

An very interesting possibility for the manifestation of thermoelectric
effects in S-N-S systems with HF metals with the weak link can be realized
in the presence of a longitudinal ultrasound. Due to the large Gr\"{u}neisen
parameter, $\Omega _{hf}\sim 10^{2}$, the longitudinal ultrasound produces
the temperature gradient oscillation with the amplitude 
\begin{equation}
\delta T\approx 2\pi \Omega _{hf}Tu_{0}/\lambda .  \label{4}
\end{equation}
Here, $\lambda $ is the wave-length of the sound and $u_{0}$ is the atomic
displacement due to the sound propagation. For the typical experimental
value \cite{Heavy3} of $\lambda \sim 10^{5}$ \AA\ and by assuming for $%
u_{0}\sim (0.1-0.01)$ \AA\ then the amplitude of the temperature oscillation
is $\delta T\sim (10^{-3}-10^{-4})$ $K$, i.e. one has $\delta T>\Delta
T_{c,hf}$. The latter inequality tells us that the longitudinal sound can
generate voltage oscillations. This effect will be elaborated elsewhere \cite
{Kulic2}.

Since the fixed value of $\Delta T$ at the S-N-S contact plays the role of
the fixed external current in S-N-S weak links, then the effects of magnetic
field, pinning centers, etc. are interrelated in these two phenomena.
Various such effects will be studied elsewhere \cite{Kulic2}.

In the above analysis it was assumed that the S-N-S system, with a HF metal
as the weak link (N), is described by the Josephson current relation. In
order to get a small value of $\Delta T_{c}$ it was also assumed that the
voltage $V_{c}(\equiv I_{c}R_{N})$ is small. These two conditions can be
realized in S-N-S systems with long spacing between superconducting banks,
i.e. $L\gg \xi _{N}=(\hbar D/2\pi T)^{1/2}$ where $\xi _{N}$ is the
coherence length of the weak link material (HF metal) and $D$ is the
diffusion coefficient. In that case one has $I_{S}R_{N}\approx (4\Delta
^{2}L/\mid e\mid T\xi _{N})\exp (-L/\xi _{N})\sin \varphi $ \cite{Likharev}.
In many applications of S-N-S weak links it is necessary that the critical
voltage $V_{c}$ is comparable with the values in Josephson tunnel junctions.
For that purpose the length of the weak link material must be short, i.e. $%
L<(2-4)\xi $ where $\xi $ is the coherence length of the weak link (HF)
material.

In conclusion we argued that the S-N-S weak links based on heavy-fermion
normal metals as the N part can be useful for thermosensors. This\ property
is due to the large thermopower of the HF metals, like for instance in $%
UPt_{3}$, which is two orders of magnitude larger than in standard alkali
metals. Due to the large Gr\"{u}neisen parameter in HF metals the
longitudinal ultrasound can generate voltage oscillation in S-N-S systems,
thus opening new possibilities for applications. Similar effects are
expected with other Kondo systems as the weak link.

{\it Acknowledgments} - The author acknowledges the support of the Deutsche
Forschungsgemeinschaft through the Forschergruppe ''Transportph\"{a}nomene
in Supraleitern und Suprafluiden''.

\end{document}